# Nonlinear polaritons in Bose- Einstein condensates in optical lattices

A.I. Maimistov, E.V. Kazantseva

Department of Solid State Physics, Moscow Engineering Physics Institute,
Kashirskoye sh. 31, Moscow, 115409, Russia*, e-mail: maimistov@mail.ru*

**ABSTRACT**

We study the interaction of a Bose-Einstein condensate in an optical lattice with additional electromagnetic fields under Raman resonance condition. System of evolution equations describing ultra-short optical pulse propagation and photo-induced transport of cold atoms in optical lattice is derived. The steady state solution of these equations was found. There are new kinds of polaritonic solitary waves propagating.

## 1. INTRODUCTION

Optical lattices are arrays of microscopic potentials induced by the interfering laser beams. This periodical potential can be used to confine cold atoms[1-5]. The dynamics of the atoms on the optical lattice realizes the Bose-Hubbard model[3,6,7] describing the hopping of bosonic atoms between the lowest energy states of the optical lattice sites. It should be pointed that the parameters of optical lattice are fully controlled by the optical methods. Thus this artificial medium provides a great potential for the new investigation of the transport both atoms and electromagnetic waves in periodic structures.

Starting point in theory is the Hamiltonian for bosonic atoms in external trapping and periodic potentials

$$H = \int d^3x \psi^+(\mathbf{r}) \left( -\frac{\hbar^2}{2m} \nabla^2 + V_0(\mathbf{r}) + V_T(\mathbf{r}) \right) \psi(\mathbf{r}) + \frac{2\pi a_s \hbar^2}{m} \int d^3x \psi^+(\mathbf{r}) \psi^+(\mathbf{r}) \psi(\mathbf{r}) \psi(\mathbf{r}).$$

where $\psi(\mathbf{r})$ is a bose field operator for atoms in a given internal atomic state. $V_T(\mathbf{r})$ is slowly varying trapping potential and $V_0(\mathbf{r})$ is the optical lattice potential.

$$V_0(\mathbf{r}) = -(1/2)\alpha_m <E^2(\mathbf{r})>,$$

where $\alpha_m$ is the polarizability of the atom. If one use laser beams of two kind of wave (for example, fundamental and second harmonic waves) with phase shift between them counterpropagating in the one direction, interference between these fields produces a standing wave two periodical potential (Fig.1). In this case the direct tunneling can be neglected. By analogy with the well-known tight-binding approximation of solid-state physics the field operator can be written as the expansion in terms of the Wannier basis. It valid for weakly overlapped condensates.

$$\psi(\mathbf{r}) = \sum_j \{a_{2j} w(\mathbf{r} - \mathbf{r}_{2j}) + b_{2j+1} w(\mathbf{r} - \mathbf{r}_{2j+1})\}$$

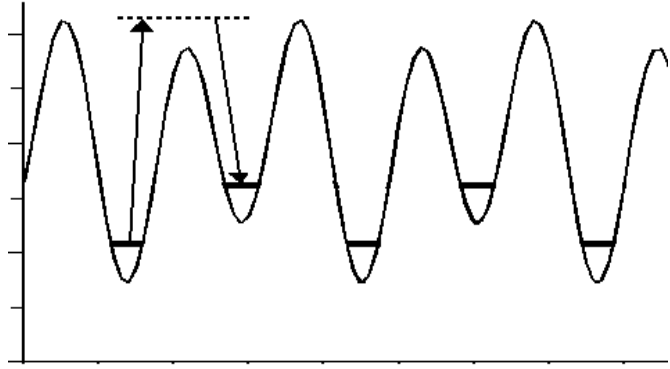

Fig.1. Atomic level scheme of two-species optical lattice

This approximation results to the two-species Bose-Hubbard model, which we will consider in this paper[3].

## 2. TWO-SPECIES BOSE-HUBBARD MODEL

Let us consider the model in where the transport of the cold atoms is governed by the photo-induced hopping in optical lattice. Like the paper[3] we will use the Hamiltonian:

$$H = -\sum_{(i,j)}(J_b a_i^+ b_j + J_a b_i^+ a_j + H.c.) + \sum_j \varepsilon_{aj} a_{2j}^+ a_{2j} + \frac{1}{2} U_{aa} \sum_j a_{2j}^{+2} a_{2j}^2 + \qquad (1)$$
$$+ \sum_j \varepsilon_{bj} b_{2j+1}^+ b_{2j+1} + \frac{1}{2} U_{bb} \sum_j b_{2j+1}^{+2} b_{2j+1}^2 + \frac{1}{2} U_{ab} \sum_{(i,j)} a_{2i}^+ a_{2i} (b_{2j+1}^+ b_{2j+1} + b_{2j-1}^+ b_{2j-1}).$$

were $a_{2j}$ and $b_{2j+1}$ are boson destruction operators referring to atoms in the internal states of the even and odd sites of optical lattice respectively. The first term in the Hamiltonian describes the Raman induced hopping between neighboring sites with couplings

$$J_{a,b} = \int d^3 x\, w_a^*(\mathbf{r}) \Omega^{(a,b)}_{eff}(\mathbf{r}) w_b(\mathbf{r}),$$

If we use the quasi-harmonic representation for electromagnetic fields

$$E_{1,2} = \mathbf{E}_{1,2} \exp(-i\omega_{1,2} t + i\mathbf{k}_{1,2} \cdot \mathbf{r}) + c.c.,$$

then the effective two-photon Rabi frequency $\Omega^{(a,b)}_{eff}(\mathbf{r})$ can be written as

$$\Omega^{(a,b)}_{eff}(\mathbf{r}) = \mu_{12} E_1 E_2^* \exp\{-i(\omega_1 - \omega_2)t + i(\mathbf{k}_1 - \mathbf{k}_2) \cdot \mathbf{r}\}$$

and the coupling constants are

$$J_a = J_0^* \exp[+i\Delta\omega_0], \quad J_b = J_0 \exp[-i\Delta\omega_0]$$

where $J_0^* = \mu E_1 E_2^*$. $\varepsilon_{a2j}$ and $\varepsilon_{b2j+1}$ describe the energies offset of each lattice site. On site interactions of atoms $a$ and $b$ are described by $U_{aa}$ and $U_{bb}$, and a nearest-neighbor interaction is described by $U_{ab}$.

## 3. EQUATIONS OF MOTION

The Heisenberg equations of motion derived from the Hamiltonian (1) are given by

$$i\hbar \frac{\partial}{\partial t} a_{2j} = -J_0^* e^{i\Delta\omega_0 t}(b_{2j+1} + b_{2j-1}) + \varepsilon_{a,2j} a_{2j} + U_{aa} a_{2j}^+ a_{2j} a_{2j} + $$
$$+ U_{ab}(b_{2j+1}^+ b_{2j+1} + b_{2j-1}^+ b_{2j-1}) a_{2j}, \quad (2)$$

$$i\hbar \frac{\partial}{\partial t} b_{2j+1} = -J_0 e^{-i\Delta\omega_0 t}(a_{2j} + a_{2j+2}) + \varepsilon_{b,2j+1} b_{2j+1} + $$
$$+ U_{bb} b_{2j+1}^+ b_{2j+1} b_{2j+1} + U_{ab}(a_{2j+2}^+ a_{2j+2} + a_{2j}^+ a_{2j}) b_{2j+1}. \quad (3)$$

where $\Delta\omega_0 = (\omega_1 - \omega_2)$. Using the coherent state representation, these equations can be rewritten in the following form

$$i\hbar \frac{\partial}{\partial t} a_{2j} = -J_0^* e^{i\Delta\omega_0 t}(b_{2j+1} + b_{2j-1}) + \varepsilon_{a,2j} a_{2j} + $$
$$+ U_{aa} |a_{2j}|^2 a_{2j} + U_{ab}(|b_{2j+1}|^2 + |b_{2j-1}|^2) a_{2j}, \quad (4)$$

$$i\hbar \frac{\partial}{\partial t} b_{2j+1} = -J_0 e^{-i\Delta\omega_0 t}(a_{2j} + a_{2j+2}) + \varepsilon_{b,2j+1} b_{2j+1} + $$
$$+ U_{bb} |b_{2j+1}|^2 b_{2j+1} + U_{ab}(|a_{2j}|^2 + |a_{2j+2}|^2) b_{2j+1}. \quad (5)$$

It is suitable to introduce new variables

$$\tilde{a}_j = a_j \exp[it(\psi - \Delta\omega_0/2)], \quad \tilde{b}_j = b_j \exp[it(\psi + \Delta\omega_0/2)],$$

with $\psi = (\varepsilon_a + \varepsilon_b)/2\hbar$..

$$i\hbar \frac{\partial}{\partial t} \tilde{a}_{2j} = -J_0^*[\tilde{b}_{2j+1} + \tilde{b}_{2j-1}] - \frac{1}{2}\Delta\varepsilon \tilde{a}_{2j} + $$
$$+ U_{aa} |\tilde{a}_{2j}|^2 \tilde{a}_{2j} + U_{ab}(|\tilde{b}_{2j+1}|^2 + |\tilde{b}_{2j-1}|^2) \tilde{a}_{2j}, \quad (6)$$

$$i\hbar \frac{\partial}{\partial t} \tilde{b}_{2j+1} = -J_0[\tilde{a}_{2j} + \tilde{a}_{2j+2}] + \frac{1}{2}\Delta\varepsilon \tilde{b}_{2j+1} + $$
$$+ U_{bb} |\tilde{b}_{2j+1}|^2 \tilde{b}_{2j+1} + U_{ab}(|\tilde{a}_{2j}|^2 + |\tilde{a}_{2j+2}|^2) \tilde{b}_{2j+1}. \quad (7)$$

where $\Delta\varepsilon = (\varepsilon_b - \varepsilon_a) - \hbar\Delta\omega_0$.

When considering an electromagnetic pulse in the quasi-harmonic approximation (in the approximation of a slowly varying envelope and phase), it may be additionally assumed that the envelope encompasses a large number of sites, so that the transition to the continuous approximation is justified; i.e., one may write

$$f_{j\pm1} \approx f(j\Delta x) \pm \Delta x \, \partial f / \partial x + (\Delta x/2) \partial^2 f / \partial x^2$$

for any discrete variables. In this situation, total system of the evolution equations can be written as

$$i\hbar \frac{\partial}{\partial t}\tilde{a} = -\frac{1}{2}\Delta\varepsilon\tilde{a} + U_{aa}|\tilde{a}|^2 \tilde{a} + U_{ab}(2|\tilde{b}|^2 + \Delta x^2 \frac{\partial^2}{\partial x^2}|\tilde{b}|^2)\tilde{a} - \\ -\mu E_1 E_2^*[2\tilde{b} + \Delta x^2 \frac{\partial^2}{\partial x^2}\tilde{b}],$$  (8)

$$i\hbar \frac{\partial}{\partial t}\tilde{b} = +\frac{1}{2}\Delta\varepsilon\tilde{b} + U_{bb}|\tilde{b}|^2 \tilde{b} + U_{ab}(2|\tilde{a}|^2 + \Delta x^2 \frac{\partial^2}{\partial x^2}|\tilde{a}|^2)\tilde{b} - \\ -\mu E_2 E_1^*[2\tilde{a} + \Delta x^2 \frac{\partial^2}{\partial x^2}\tilde{a}].$$  (9)

$$\left(\frac{\partial}{\partial x} + \frac{1}{V_1}\frac{\partial}{\partial t}\right)E_1 = +2\pi i k_1 n_A \mu E_2 \tilde{a}^*[2\tilde{b} + \Delta x^2 \frac{\partial^2}{\partial x^2}\tilde{b}],$$  (10)

$$\left(\frac{\partial}{\partial x} + \frac{1}{V_2}\frac{\partial}{\partial t}\right)E_2 = +2\pi i k_2 n_A \mu E_1 \tilde{b}^*[2\tilde{a} + \Delta x^2 \frac{\partial^2}{\partial x^2}\tilde{a}].$$  (11)

Here $k_{1,2}$ is wave vector, $V_{1,2}$ is group velocity of light pulse corresponding to frequency of carry wave $\omega_{1,2}$. $n_A$ is the density of atoms in a trap. In this situation, the $U_{ab}$ terms describe a correction to the Lorentz local field and the terms containing derivative of second order account for the motion of an exciton along the optical lattice. The effective mass of such an exciton is defined by the coupling constant $J_0$. There is an *excitonic wave*, i.e., a wave transferring an excited state (an exciton) from one site to another. The velocity of this wave is governed by the exchange-interaction constant $J_0$ and can be much smaller than the velocity of the electromagnetic wave. A *polaritonic wave* is an electromagnetic wave propagating along the optical lattice with a velocity differing only slightly from the light velocity owing to the drag of the polarization cloud. In the approximation of heavy excitons, the derivatives of second order in these equations should be disregarded. The equations obtained in this approximation become local (all variables depend on the same coordinate). The remaining equations can now be written in terms of real variables as

$$\frac{\partial}{\partial t}v = -\left(\frac{\partial \Phi}{\partial t} - \frac{\Delta}{\hbar} + \frac{\Delta U N_0}{\hbar}n\right)u, \qquad \frac{\partial}{\partial t}n = -\frac{4\mu}{\hbar}A_1 A_2 u$$  (12.1)

$$\frac{\partial}{\partial t}u = -\left(\frac{\partial \Phi}{\partial t} - \frac{\Delta}{\hbar} + \frac{\Delta U N_0}{\hbar}n\right)v + \frac{4\mu}{\hbar}A_1 A_2 n,$$  (12.2)

that are the generalized Bloch equations, and

$$\left(\frac{\partial}{\partial x}+\frac{1}{V_1}\frac{\partial}{\partial t}\right)A_1 = -2\pi k_1 n_A \mu A_2 u, \quad \left(\frac{\partial}{\partial x}+\frac{1}{V_2}\frac{\partial}{\partial t}\right)A_2 = 2\pi k_2 n_A \mu A_1 u, \tag{13.1}$$

$$A_1\left(\frac{\partial}{\partial x}+\frac{1}{V_1}\frac{\partial}{\partial t}\right)\varphi_1 = 2\pi k_1 n_A \mu A_2 v, \quad A_2\left(\frac{\partial}{\partial x}+\frac{1}{V_2}\frac{\partial}{\partial t}\right)\varphi_2 = 2\pi k_2 n_A \mu A_1 v. \tag{13.2}$$

that are reduced Maxwell equations. Here the next real variables and parameters are employed

$$\sigma = 2\tilde{a}\tilde{b}^* = N_0(v-iu)\exp(i\Phi), \quad n = N_o(|\tilde{a}|^2 - |\tilde{b}|^2),$$
$$E_{1,2} = A_{1,2}\exp(i\varphi_{1,2}), \quad \Phi = \varphi_2 - \varphi_1,$$
$$\Delta = \Delta\varepsilon - N_0(U_{aa} - U_{bb}), \quad \Delta U = (U_{aa}+U_{bb})/2 - 2U_{ab}.$$

where $N_0$ is number of atoms in the site of optical lattice.

### 4. SOME EXAMPLES OF STEADY STATE WAVES

Let consider the pulse propagation of solitary polaritonic waves as traveling waves with a non-varying profile. In order to get these steady state solutions, one should assume that all field and atomic variables depend on a single variable $\xi = \omega_0(t-x/V)$, where $V$ is a velocity of this wave. Henceforth we will consider the case of normal dispersion, i.e. $V_1 > V_2$.

From (13) one can obtain the integral of motion

$$\frac{V_1-V}{k_1 V_1}A_1^2 + \frac{V_2-V}{k_2 V_2}A_2^2 = const, \tag{14}$$

The second invariant results from the Bloch equations (12):

$$n^2 + v^2 + u^2 = 1. \tag{15}$$

Let us consider the following initial and boundary conditions

$$A_{1,2}(t=0,x) = 0, \text{ and } u = v = 0 \text{ at } x \to \pm\infty. \tag{19}$$

Hence *const* =0. The expression (15) is valid only if $V_1 > V > V_2$. Thus, one has

$$A_1^2 = k_1 V_1 (V-V_2)/k_2 V_2 (V_1-V) A_2^2.$$

The steady state solutions of the system of equations (12)-(13) are existed only under condition of the preliminary frequency detuning: $\Delta\varepsilon - N_0(U_{aa}-U_{bb}) = N_0\Delta U = N_0[(U_{aa}+U_{bb})/2 - 2U_{ab}]$. From equations (12)-(13) it follows

$$n = 1 - q/\kappa, \quad v = (\beta/2\kappa^2)q,$$

where $q$ was defined by formula $q = (4\mu/\hbar\omega_0)(k_2 V_2 (V_1 - V)/k_1 V_1 (V - V_2))^{1/2} A_1^2$.

This solution can be written for normalized envelope of the light pulse in following form

$$q = \frac{2\kappa}{(1+\beta^2/4\kappa^2)+\kappa^2\xi^2}, \qquad (20)$$

where we introduce the parameters

$$\beta = \Delta U N_0 / \hbar\omega_0, \quad \kappa = 2\pi n_A \mu V / \omega_0 \left(k_1 V_1 k_2 V_2 /(V_1 - V)(V - V_2)\right)^{1/2}$$

One may name this pulse "algebraic soliton", due to its decay rate with time and coordinate.

Now let us consider the following initial and boundary conditions

$$A_1(t=0, x) = A_0, \quad A_2(t=0, x) = 0 \text{ and } u = v = 0 \text{ at } x \to \pm\infty.$$

The integral of motion takes the form

$$\frac{V_1 - V}{k_1 V_1} A_1^2 + \frac{V_2 - V}{k_2 V_2} A_2^2 = \frac{V_1 - V}{k_1 V_1} A_0^2$$

We will consider the case of $V > V_1 > V_2$ (i.e., it is a fast polaritonic wave). After introducing the new dependent vartiables

$$A_1 = A_0 q_1, \quad A_2 = (k_2 V_2 (V - V_1)/k_1 V_1 (V - V_2))^{1/2} A_0 q_2,$$

and parameters

$$\omega_0 = 4\mu A_0^2/\hbar \left(k_2 V_2 (V - V_1)/k_1 V_1 (V - V_2)\right)^{1/2}, \quad \kappa = 2\pi n_A \mu V / \omega_0 \left(k_1 V_1 k_2 V_2 /(V - V_1)(V - V_2)\right)^{1/2},$$

the system of equations (12)-(13) takes the following form

$$\frac{\partial}{\partial \xi} v = -\left(\frac{\partial \Phi}{\partial \xi} - \delta + \beta n\right) u, \qquad \frac{\partial}{\partial \xi} n = -q_1 q_2 u$$

$$\frac{\partial}{\partial \xi} u = -\left(\frac{\partial \Phi}{\partial \xi} - \delta + \beta n\right) v + q_1 q_2 n, \qquad (21)$$

$$\frac{\partial}{\partial \xi} q_1 = -\kappa q_2 u, \quad \frac{\partial}{\partial \xi} q_2 = \kappa q_1 u, \quad \frac{\partial}{\partial \xi} \Phi = \kappa v \frac{q_1^2 - q_2^2}{q_1 q_2}$$

The steady state solutions of this system of equations are existed if the preliminary frequency detuning $\delta = \beta - (\beta/2\kappa)$ is provided. From these equation one can find that

$$q_1^2 + q_2^2 = 1, \quad n = 1 - q_2^2/2\kappa, \quad u = (\partial q_2/\partial \xi)/\kappa\sqrt{1-q_2^2}. \qquad (22)$$

Taking into account these expressions and using the other equations, one can obtain the following equation

$$\frac{\partial v}{\partial \xi} = -v\left(\frac{1-2q_2^2}{q_2(1-q_2^2)}\right)\frac{\partial q_2}{\partial \xi} + \left(\frac{\delta-\beta+(\beta/2\kappa)q_2^2}{\kappa\sqrt{1-q_2^2}}\right)\frac{\partial q_2}{\partial \xi}.$$

The nonsingular solution of this equation is

$$v(q) = -\left(\frac{\beta}{8\kappa^2}\right)\sqrt{1-q_2^2}. \qquad (23)$$

The substitution of the (22) and (23) into (15) leads to

$$\frac{4}{1-q_2^2}\left(\frac{\partial q_2}{\partial \xi}\right)^2 = \left(4\kappa - \left(\frac{\beta}{4\kappa}\right)^2\right)q^2 - \left(1 - \left(\frac{\beta}{4\kappa}\right)^2\right)q^4$$

This equation has the following solutions, which corresponds to solitary wave (Fig.2)

$$q_2^2(\xi) = \frac{4\kappa - m^2}{1 - m^2 + (4\kappa - 1)\cosh^2 0.5\xi\sqrt{4\kappa - m^2}} \qquad (24)$$

where $m = \beta/4\kappa$ represent of the measure of the interatomic interaction in site of optical lattice.

The limit $m^2 \to 4\kappa$ results to algebraic soliton

$$q_2^2(\xi) = \frac{1}{1+(4\kappa-1)\xi^2/2}.$$

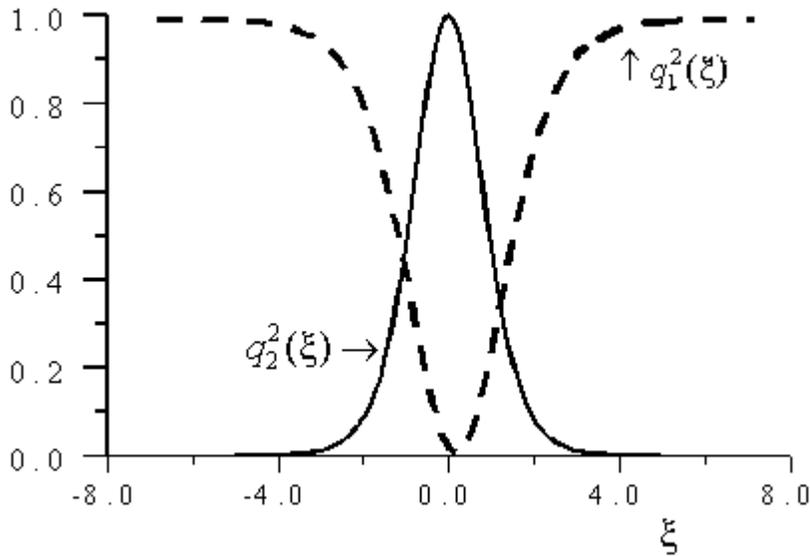

*Fig.2*. Envelope of the bound pair of bright and dark pulses

By considering of the another boundary conditions we should find the periodical solutions of different kinds.

## 4. CONCLUSION

We have introduced and analyzed a model for the propagation of ultra-short pulses of the electromagnetic field in a optical lattice containing cold bosonic atoms. In frame of this model new system of the generalized Maxwell-Bloch equations was derived. It was found a different kind of steady state solitary waves that represent the propagation of electromagnetic wave coupled with atomic one. Among these waves there is bound pair of bright and dark pulses.

## ACKNOWLEDGMENTS

We are grateful to Dr. S.O. Elyutin for enlightening discussions. This research was supported in part by Russian Foundation of Basic Research (grant № 03-02-16979).

## REFERENCES

1. W. Greenwood, P. Pax, and P. Meystre, "Atomic transport on one-dimensional optical lattices", *Phys.Rev*. **A 56**, 2109-2122 (1997).
2. P.M.Visser and G.Nienhuis "Quantum transport of atoms in an optical lattice", *Phys.Rev*. **A56**, 3950-3961 (1997).
3. D. Jaksch, C. Bruder, J. I. Cirac, C. W. Gardiner, and P. Zoller "Cold Bosonic Atoms in Optical Lattices", *Phys.Rev.Lett*. **81**, 3108-3111 (1998).
4. F.Kh. Abdullaev, B.B. Baizakov, S.A. Darmanyan, V.V. Konotop, and M. Salerno, "Nonlinear excitations in arrays of Bose-Einstein cindensates", *Phys.Rev*. **A64**, 043606 (2001).
5. G. Grynberg and C. Robilliard "Cold atoms in dissipative optical lattices", *Phys.Reports*, **355,** 335-451 (2001).
6. C. Bruder, Rosario Fazio, Gerd Schon, "Superconductor-Mott-insulator transition in Bose systems with finite-range interactions", Phys.Rev. **B47**, 342-347 (1992).
7. R. Roth and K. Burnett, "Superfluidity and interference pattern of ultracold bosons in optical lattices", *Phys.Rev*. **A67**, 031602(R) (2003).